\documentclass[prl, twocolumn, superscriptaddress, showpacs,amsmath,amssymb]{revtex4}

\usepackage{float}
\usepackage{graphicx}
\usepackage{subfigure}

\newcommand{\ts}[2]{{#1}_{\text{#2}}}

\begin{document}

\title{Coherent population trapping of electron spins in a high-purity n-type GaAs Semiconductor}
\author{Kai-Mei C. Fu}
\email{kaimeifu@stanford.edu}
\affiliation{Quantum Entanglement
Project, ICORP, JST, Edward L.
    Ginzton Laboratory, Stanford University, Stanford, California
    94305-4085, USA}
\author{Charles Santori}
\affiliation{Quantum Science Research, Hewlett-Packard
Laboratories, 1501 Page Mill Road, MS1123, Palo Alto, California
94304}
\author{Colin Stanley}

\author{M.C. Holland}
\affiliation{Department of Electronics and Electrical Engineering,
Oakfield Ave, University of Glasgow, Glasgow, G12 8LT, U.K.}
\author{Yoshihisa Yamamoto}
\altaffiliation[Also at ]{National Institute of Informatics,
Tokyo, Japan} \affiliation{Quantum Entanglement Project, ICORP,
JST, Edward L.
    Ginzton Laboratory, Stanford University, Stanford, California
    94305-4085, USA}

\begin{abstract}

In high-purity $n$-type GaAs under strong magnetic field, we are
able to isolate a lambda system composed of two Zeeman states of
neutral-donor bound electrons and the lowest Zeeman state of bound
excitons. When the two-photon detuning of this system is zero, we
observe a pronounced dip in the excited-state photoluminescence
indicating the creation of the coherent population-trapped state.
Our data are consistent with a steady-state three-level
density-matrix model. The observation of coherent population
trapping in GaAs indicates that this and similar semiconductor
systems could be used for various EIT-type experiments.

\end{abstract}

\maketitle

In the past decade great steps have been made toward the coherent
control of light using techniques based on electromagnetically
induced transparency (EIT)~\cite{Harris97a}.  Light has been
slowed by seven orders of magnitude~\cite{Hau99a}, stored and
released on command~\cite{Fleischhauer00a, Liu01a, Turukhin02a},
and coherently manipulated while stored in atomic
states~\cite{Mair02a,Bajcsy03a}.  The applications of an
integrated EIT system for quantum information processing are
numerous: robust entanglement creation for quantum
repeaters~\cite{Duan01a}, single photon detection~\cite{Harris98a}
and single photon-storage~\cite{Liu01a, Phillips01a} for linear
optics quantum computation~\cite{Knill01a}, and the creation of
large optical non-linearities~\cite{Schmidt96a} for photonic gates
in non-linear optics quantum computation~\cite{Chuang95a,
Barrett04a}.

%
%
\begin{figure}
\subfigure[] {\includegraphics[height =30mm,
keepaspectratio]{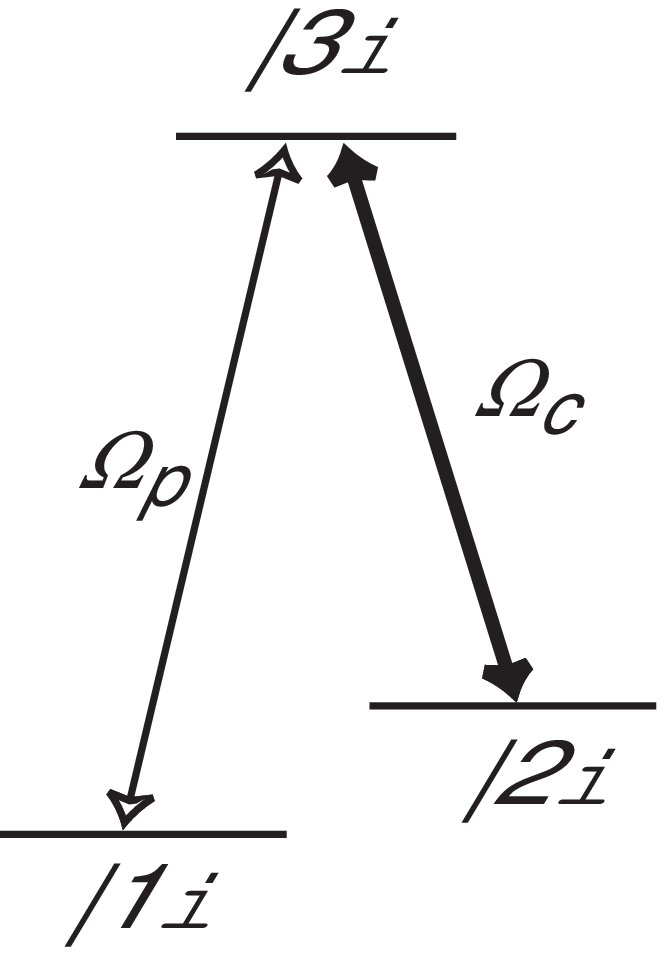}} \quad \subfigure[]{
\includegraphics[height = 30mm,
keepaspectratio]{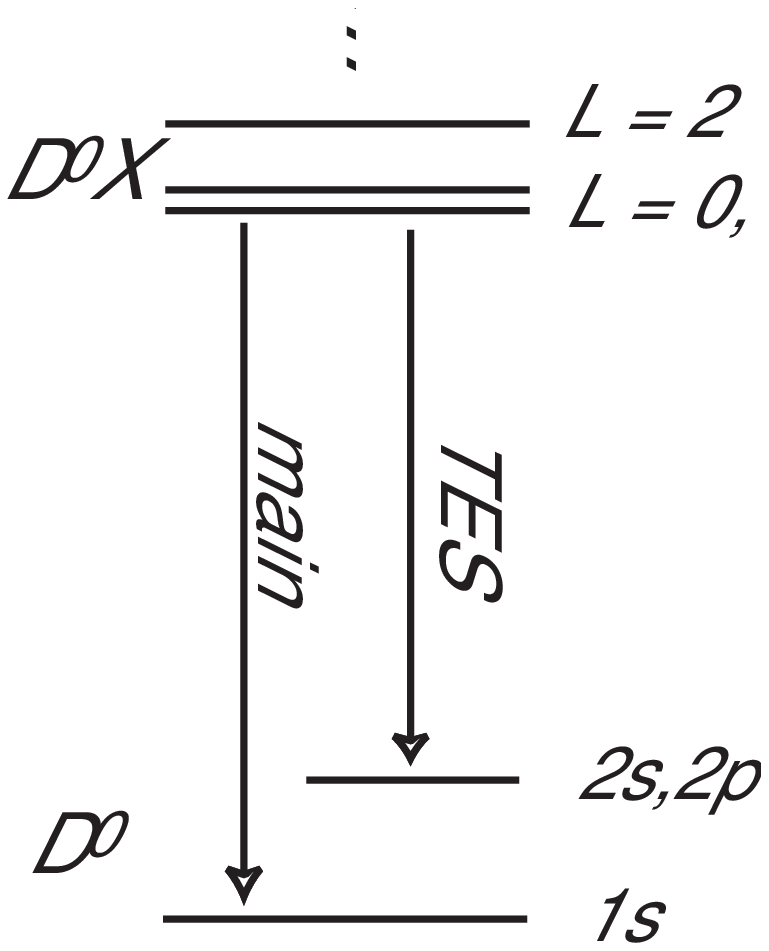}} \quad
\subfigure[]{\includegraphics[height = 30mm,
keepaspectratio]{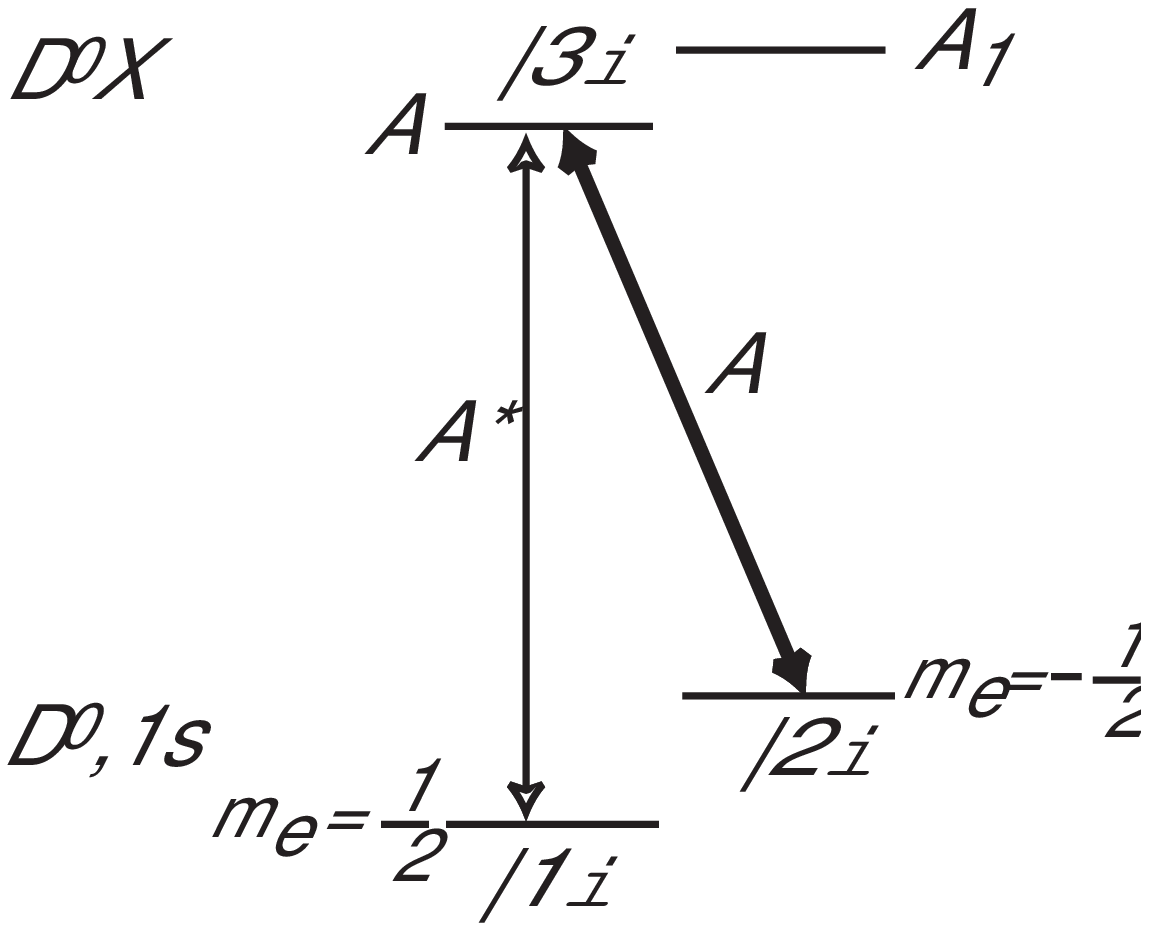}} \caption{(a) 3-level lambda system.
(b) Energy level diagram of the D$^0$X and D$^0$ states at 0~T. In
all experiments we resonantly excite the main transitions and
probe the excited-state population via the TES photoluminescence.
(c) Energy level diagram of D$^0$ 1s and lowest two D$^0$X states
in an applied magnetic field. A complete energy diagram of the
GaAs D$^0$X states in a magnetic field is found in
Ref.~\cite{Karasyuk94a}. In the spherical approximation, state A
corresponds to hole state L = 0, $m_h = -\frac{3}{2}$ and state
A$_1$ corresponds to L = 1, $m_h = -\frac{1}{2}$. Transition A$^*$
is allowed due to the lack of spherical approximation in the
crystal lattice.  To create a population-trapped state we apply a
strong coupling field to transition A and a weak probe field to
A$^*$.} \label{energyfig}
\end{figure}
%
%

EIT is based on the effect of coherent population trapping which
was first observed in the 1970's in atomic gases
~\cite{Alzetta76a, Gray78a}. In a three-level $\Lambda$-system, a
probe field with Rabi frequency $\ts\Omega{p}$ couples states
$|1\rangle$ and $|3\rangle$, and a coupling field with Rabi
frequency $\ts\Omega{c}$ couples states $|2\rangle$ and
$|3\rangle$ (Fig.~\ref{energyfig}a). Optical pumping leads to a
coherent superposition of states $|1\rangle$ and $|2\rangle$ that
is decoupled from $|3\rangle$ due to a quantum interference
between the two transitions. The crucial condition for coherent
population trapping is that the decoherence rate $\gamma_{12}$
between states $|1\rangle$ and $|2\rangle$ is slow compared to the
radiative decay rate of $|3\rangle$. Furthermore, for
photon-storage applications, $\gamma_{12}^{-1}$ determines how
long quantum information can be stored~\cite{Fleischhauer00a}.

Long decoherence times, which naturally arise in atomic systems
\cite{Liu01a}, are also possible in solids~\cite{Hemmer01a,
Tyryshkin03a, Kennedy03a}. EIT has been observed in rare-earth
doped insulators~\cite{Turukhin02a}, N-V centers in
diamond~\cite{Hemmer01a}, and in the transient optical response of
GaAs quantum wells~\cite{Phillips02a}. Here, we consider electron
spins bound to neutral donors (D$^0$) in a semiconductor, a system
that could offer some unique advantages. For example, the optical
transitions to the donor-bound exciton states feature a small
inhomogeneous broadening (2~GHz) combined with a large oscillator
strength (1~ns radiative lifetime~\cite{Hwang73a}). Furthermore,
the ground state is long-lived, unlike the exciton states used in
previous semiconductor EIT experiments~\cite{Phillips02a}.
Finally, donor impurities can easily be integrated into monolithic
microcavities. In this letter, we report the observation of
coherent population trapping in an ensemble of D$^0$ spins,
demonstrating that a lambda system can be optically addressed and
manipulated.  While the degree of ground-state coherence currently
obtainable is small, it is thought to be limited mainly by
inhomogeneous broadening of the electron-Zeeman splitting, which
can hopefully be remedied in pulsed experiments with spin-echo
techniques.

The energy level structure of a neutral donor is shown in
Fig.~\ref{energyfig}b.  Due to the small electron effective mass
and high dielectric constant of GaAs, the wavefunction of a
neutral-donor bound electron (D$^0$) extends over many lattice
sites and is well described by the hydrogenic wavefunction with a
100~\AA~Bohr radius~\cite{Skromme85a}. With an applied magnetic
field, the 1s state splits into the two electron-Zeeman spin
states which are labelled $|1\rangle$ and $|2\rangle$ in
Fig.~\ref{energyfig}c. The excited states consist of an
electron-hole pair, or exciton, bound to the D$^0$ center. This
donor bound-exciton complex (D$^0$X), consisting of two electrons
in a spin-singlet state, a hole with quasi-spin-3/2, and the donor
impurity, can be resonantly excited from the D$^0$ state.  At zero
magnetic field, the D$^0$X is composed of closely spaced orbital
angular momentum states (Fig.~\ref{energyfig}b). In a magnetic
field each of the D$^0$X states splits into the four hole-Zeeman
spin states. In Fig.~\ref{energyfig}c we identify the
lowest-energy D$^0$X states as A and A$_1$ following
Ref.~\cite{Karasyuk94a}. We denote transitions to the D$^0$ state
$|$m$_e= -\frac{1}{2}\rangle$ with a label only (e.g.~A) and
transitions to the state $|$m$_e = \frac{1}{2}\rangle$ with an
asterisk (e.g.~A$^*$). Although the D$^0$X predominately relaxes
to the D$^0$ 1s state, there is a small probability it will decay
to an excited orbital D$^0$ state. These transitions are called
`two electron satellites' or TES (Fig.~\ref{energyfig}b).

\begin{figure}
{\includegraphics[height = 37 mm,
keepaspectratio]{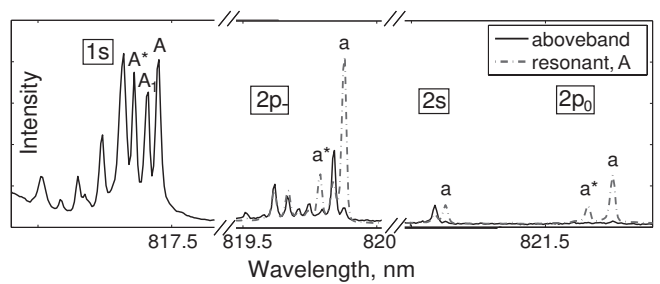}}\caption{Above-band and resonant
excitation photoluminescence spectra of D$^0$X.  Transitions A and
A$^*$ are clearly resolved.  With resonant excitation of
transition A, the TES lines associated with state A (labelled $a$,
$a^*$) are noticeably enhanced.  A detailed assignment of the
D$^0$X excited states can be found in Ref.~\cite{Karasyuk94a}.
Above-band TES intensities are 10 times the actual intensity.
Resolution is spectrometer limited. } \label{spectrafig}
\end{figure}

Our sample consisted of a 10~$\mu$m GaAs layer on a 4~$\mu$m
Al$_{0.3}$Ga$_{0.7}$As layer grown by molecular-beam epitaxy on a
GaAs substrate. The sample had a donor concentration of
$\sim$5$\times$10$^{13}$~cm$^{-3}$. We mounted the sample
strain-free in a magnetic cryostat in the Voigt ($\vec{k} \perp
\vec{B}$) geometry. A photoluminescence (PL) spectrum of the
D$^0$X emission at 7~T and 1.5~K is shown in
Fig.~\ref{spectrafig}.  With above-band excitation, the A and
A$^*$ transitions are clearly resolved.  In addition, we can
identify the TES lines associated with state A by resonantly
exciting the A or A$^*$ transitions and observing enhancements in
the associated TES lines.

%
%
%
%

Coherent population trapping can be observed as a decrease in the
excited-state population when two-photon resonance occurs.  In our
experiment, we monitor the excited state population using the TES
fluorescence.  In photoluminescence excitation (PLE) scans, an
external-cavity diode laser resonant with the A transition
(817.448~nm) (see Fig.~\ref{energyfig}c) provides the `coupling'
field, and a ring Ti:Sapphire laser, scanned across the A*
transition (817.358~nm), provides the `probe' field. The scan
resolution was measured using an optical spectrum analyzer to be
better than 10~MHz. The energy splitting between the two
transitions corresponds to the 7~T electron-Zeeman energy. The
observed g-factor $|g| = 0.41$ is close to the previously measured
$g = -0.43$ determined by the 2$p_-$ splitting~\cite{Karasyuk94a}.

\begin{figure}
{\includegraphics[height =75mm,
keepaspectratio]{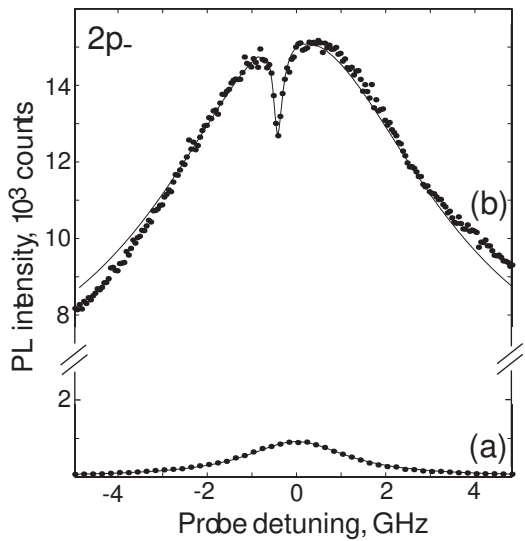}}\caption{(a) Probe only PLE scan over
A$^*$ transition.  PL is collected from the ``a'' TES line at
2$p_-$. Probe laser intensity is $\sim$~0.15~W/cm$^2$. (b) PLE
scan over the A$^*$ transition with the coupling laser resonant on
the A transition. Coupling intensity is $\sim$~2.5~W/cm$^2$. A dip
in the PLE intensity at zero two-photon detuning due to coherent
population trapping is observed. Identical behavior is also
observed for the 2$p_0$ and 2$s$ TES lines. The solid line is a
theoretical fit by the three-level density-matrix model described
in the text. The fitting parameters are $\ts\Omega{c}$ = 650~MHz,
$\ts\Omega{p}$ = 16~MHz, $\Gamma_{21}$ = (2.6~$\mu$s)$^{-1}$,
$\gamma_{3a}$ = 4.6~GHz, $\gamma_{3b}$ = 22~GHz, $\Gamma_{32}$ =
1~ns$^{-1}$, $\Gamma_{31} = 0.08\Gamma_{32}$, $\gamma_{12}$ =
(1.7~ns)$^{-1}$.} \label{EITfig}
\end{figure}

The results from a representative scan are shown in
Fig.~\ref{EITfig}.  We discuss three scenarios:  probe-laser only
excitation, two laser excitation detuned from resonance, and
two-laser excitation on two-photon resonance. With the probe laser
only, the PLE spectrum gives a linewidth of only 2~GHz. The data
fit a Lorentzian lineshape extremely well and indicate that there
is little inhomogeneous broadening.  The emission intensity is
weak due to optical pumping of most of the electron population
into state $|2\rangle$, which for the probe laser only, is a dark
state.  With both lasers exciting the sample but detuned from
two-photon resonance, the emission becomes much stronger since in
this case there is no dark state.  When the probe and coupling
lasers are brought into two-photon resonance, a pronounced and
narrow reduction of the emission intensity is observed as a new
dark state is formed which is a coherent superposition of states
$|1\rangle$ and $|2\rangle$.

The decrease in the excited-state population observed on
two-photon resonance is incomplete because of decoherence and
population relaxation between levels $|1\rangle$ and $|2\rangle$.
The results can be understood in terms of a 3-level system
interacting with a reservoir, described by the density-matrix
master equation:
$$
{\frac{\partial}{\partial t}\rho = \frac{1}{i\hbar}[{H}, \rho] -
\mathcal{L}(\rho)} = 0
$$
in which $H$ is the Hamiltonian of the system and
$\mathcal{L}(\rho)$ is the Linbladian operator describing the
decoherence processes. In the interaction picture and rotating
wave approximation,
$$
H = -\hbar\left(
\begin{matrix}
0 & 0 & \frac{\ts\Omega{p}^*}{2} \\ 0 & \delta & \frac{\ts\Omega{c}^*}{2} \\
\frac{\ts\Omega{p}}{2} & \frac{\ts\Omega{c}}{2} & \Delta
\end{matrix}\right)
$$
in which $\Delta$ is the probe detuning and $\delta$ is the
two-photon detuning from the electron-Zeeman splitting. The
relaxation operator $\mathcal{L}(\rho)$ is given by
\begin{widetext}
\begin{equation}
 \mathcal{L}(\rho) = \left( \begin{matrix}
 -\Gamma_{12}\rho_{11}+\Gamma_{21}\rho_{22} +
 \Gamma_{31}\rho_{33}  &

-(\frac{\Gamma_{12} + \Gamma_{21}}{2} +  \gamma_{2}) \rho_{12} &

-(\frac{\Gamma_{12} + \Gamma_{31} + \Gamma_{32}}{2} + \gamma_{3a,3b})\rho_{13} \\

-(\frac{\Gamma_{12} + \Gamma_{21}}{2} +  \gamma_{2}) \rho_{21} &

\Gamma_{12}\rho_{11} - \Gamma_{21}\rho_{22} + \Gamma_{32}\rho_{33}
&

-(\frac{\Gamma_{21} + \Gamma_{31} + \Gamma_{32}}{2} + \gamma_{3a,3b})\rho_{23} \\

-(\frac{\Gamma_{12} + \Gamma_{31} + \Gamma_{32}}{2} +
\gamma_{3a,3b})\rho_{31} &

-(\frac{\Gamma_{21} + \Gamma_{31} + \Gamma_{32}}{2} +
\gamma_{3a,3b})\rho_{32} &

-(\Gamma_{31} + \Gamma_{32})\rho_{33}

\end{matrix} \right)
\label{relaxeq}
\end{equation}
\end{widetext}
 in which $\Gamma_{12}$ ($\Gamma_{21} =
\Gamma_{12}e^\frac{E_{12}}{kT}$) is the longitudinal relaxation
rate from $|1\rangle\rightarrow|2\rangle$
($|2\rangle\rightarrow|1\rangle$), $\Gamma_{31}$ ($\Gamma_{32}$)
is the radiative relaxation from $|3\rangle\rightarrow|1\rangle$
($|3\rangle\rightarrow|2\rangle$), $\gamma_2$ is the transverse
relaxation rate between $|1\rangle$ and $|2\rangle$, and
$\gamma_{3a}$ ($\gamma_{3b}$) is the level $|3\rangle$ dephasing
without (with) the coupling field.  With these definitions, the
total lower-level decoherence rate is given by $\gamma_{12} =
\frac{1}{2}(\Gamma_{12} + \Gamma_{21}) + \gamma_2$.

%
%
%
%

Fitting $\rho_{33}$ from the above model to the measured PLE curve
gives reasonable agreement, as shown in Fig.~\ref{EITfig}. The
only parameter that must be changed to fit simultaneously both the
single laser (a) and two laser (b) scans is the level $|3\rangle$
dephasing rate. The fit indicates slow ($\mu$s) electron
population relaxation rates and fast (1-2~ns) electron decoherence
rates in our system.  Thus, the system exhibits a lower level
dephasing rate on the same order as the excited state radiative
lifetime (1~ns).  From this fit we can also obtain the ratio of
the two-state coherence, $\rho_{12}$, to the ideal case,
$\rho_{12,\text{ideal}} =
\frac{\ts\Omega{p}\ts\Omega{c}}{\ts\Omega{p}^2 + \ts\Omega{c}^2}$,
and find that $\rho_{12}$/$\rho_{12,\text{ideal}}$ = 0.23. In the
weak probe limit ($\ts\Omega{p} << \ts\Omega{c}, \Gamma_{31},
\Gamma_{32}$) $\rho_{12}$ reduces to
$$
\rho_{12} \approx
\frac{\ts\Omega{p}\ts\Omega{c}}{4\gamma_{13}\gamma_{12}+\ts\Omega{c}^2}
$$
in which $\gamma_{13}$ ($\gamma_{12}$) is the total decay rate for
$\rho_{13}$ ($\rho_{12}$) given in Eq.~\ref{relaxeq}.  From this
relation it is evident that the coherence of this system is
currently limited by the short lower-level decoherence time as
well as additional dephasing of state $|3\rangle$.

\begin{figure}
{\includegraphics[height =60mm,
keepaspectratio]{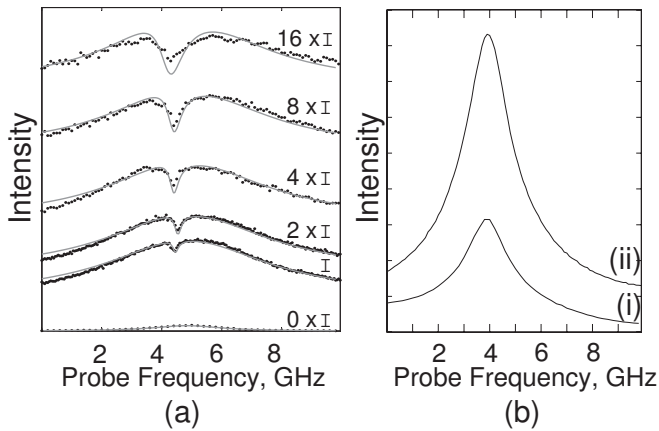}}\caption{(a)PLE scans with varying
coupling field intensities.  $I\sim$1~W/cm$^2$.  The dip becomes
wider and deeper relative to the wings of the curve as the
coupling field is increased.  The only fitting parameter varied
between the 5 two laser scans is $T$ ($\Gamma_{12} \propto
(e^{\frac{\Delta E}{kT}}-1)^{-1}$). In order of increasing
coupling intensity, $T$~=1.5, 1.5, 2, 3.1, 4.6, 6.0~K. (b)
Incoherent optical pumping experiment with (i) coupling laser off,
(ii) coupling laser on. The coupling laser is tuned to transition
A. The probe laser is scanned over A$_1^*$ (see
Fig~\ref{energyfig}c). An enhancement of the PLE intensity in the
two laser case is observed without a dip at zero detuning. }
\label{seriesfig}
\end{figure}

Additional measurements to verify the theoretically expected
behavior are shown in Fig.~\ref{seriesfig}. In
Fig.~\ref{seriesfig}a, PLE scans were performed at several
coupling intensities. As the coupling intensity increases, the
population-trapped window at zero two-photon detuning becomes
relatively wider and deeper as expected. The data fit our
theoretical model if the lower-level population-relaxation rate is
allowed to increase with increased coupling field intensity. This
increase could be due to sample heating at large coupling-laser
powers. In our sample, the GaAs substrate was not removed and
absorbs all of the incident radiation. If we assume a one-phonon
spin-orbit relaxation process~\cite{Khaetskii01a,Woods02a}, we are
able to simultaneously fit the coupling power dependence series by
varying only the sample temperature from 1.5~K to 6~K. In a second
experiment (Fig.~\ref{seriesfig}b), the two lasers are tuned to
different excited states and the PLE dip is not observed. In this
case, the probe laser is tuned to the A$_1^*$ transition and the
coupling laser is tuned to the A transition (see
Fig.~\ref{energyfig}c). As in the previous case, if only the probe
laser is applied, population becomes depleted from state
$|1\rangle$ and the PLE intensity is weak. The coupling laser
repopulates this state and the PLE intensity is enhanced. The
absence of the dip in this experiment as well as the narrow dip in
the lambda system (FWHM $\ll$ homogeneous broadening) indicate
that our results cannot be explained by standard spectral hole
burning.

%

We only observe a modest suppression of the excited-state
population.  This is due to the 1-2~ns inhomogeneous decoherence
time, T$_2^*$.  At the extremely low densities
($\sim5\times10^{13}$~cm$^{-3}$) in our sample, the
nuclear-electron hyperfine interaction becomes very
efficient~\cite{Dzhioev02a}. At these densities, the donor
electrons are well localized and do not interact with each other.
At 2~K the nuclei are essentially unpolarized and theoretical
calculations predict nanosecond T$_2^*$ due to the random nuclear
states~\cite{Merkulov02a}. Experimentally, a 5~ns T$_2^*$ has been
measured in $n$-type GaAs with $n\sim3\times10^{14}$~cm$^{-3}$ via
optically detected electron-spin resonance~\cite{Colton04a}. This
result is consistent with our value given our sample's lower donor
density. Additionally, we find that if we increase the temperature
of our sample up to 6~K, although the overall PLE intensity
decreases dramatically, the width of the dip does not change
significantly. This indicates that T$_2^*$ in our system is not
temperature dependent and is consistent with the nuclear-electron
hyperfine decoherence model.

Although the \emph{inhomogeneous} T$_2^*$ limits the depth of the
population-trapped dip, in an EIT-type experiment with pulsed
lasers and electron spin-echo techniques, the storage time should
be limited by the \emph{homogeneous} decoherence time, T$_2$.
T$_2$ of electron spins in GaAs has not been measured but could be
close to the population relaxation time, on the order of
microseconds. It has also been proposed that further improvements
on storage time could be made by transferring the electron spin
coherence to the nuclear spins. If this is achieved, a storage
time on the order of seconds may be feasible \cite{Taylor03a}.

In summary, we have observed coherent population trapping of
donor-bound electrons in GaAs.  To our knowledge, this is the
first demonstration of a lambda system in a semiconductor that
utilizes the true electron ground states. In addition, due to the
substitutional nature of donor impurities and high crystal
quality, this system has little inhomogeneous broadening in the
optical transitions. Although current population trapping is
limited by a short T$_2^*$, there exist several possible ways to
engineer this system for long T$_2$ and storage times.  Spin echo
techniques and electron to nuclear information transfer should be
able to extend possible storage times by orders of magnitude in
GaAs. Additionally, the D$^0$X system exists in every
semiconductor. Thus, a crystal composed of nuclear spin-0 elements
would significantly extend the storage lifetime.  We also note
that larger bandgap semiconductors have larger effective masses,
larger D$^0$ binding energies, and thus larger D$^0$X binding
energies \cite{Haynes60a}, which allow higher temperature
operation.

Financial support was provided by the MURI Center for photonic
quantum information systems (ARO/ARDA Program DAAD19-03-1-0199)
and NTT Basic Research Laboratories. This material was based upon
work under a National Science Foundation Graduate Fellowship. We
would like to thank T.D. Ladd, J.R. Goldman, K. Sanaka, and S.
Clark for their help with the experimental aspects of this work.
We also would like to thank S.E. Harris, D. Gammon, S. Yelin and
M.D. Lukin for the valuable discussions.


\end{document}